\documentclass[a4paper,11pt]{article}
\usepackage{pos,lineno}
\newcommand*{\ttbar}{\mathrm{t}\overline{\mathrm{t}}}
\title{Measurements of top quark properties at the LHC}
\author*[a,b]{Efe Yazgan}
\affiliation[a]{on behalf of the ATLAS and CMS Collaborations}

\affiliation[b]{National Taiwan University,\\
  Department of Physics, Laboratory of High Energy Physics, 10617 Taipei, Taiwan}

\emailAdd{efe.yazgan@cern.ch}
\abstract{Recent measurements of top quark properties at the LHC made with the ATLAS and CMS experiments are discussed. The presented results include top quark mass, width, top quark Yukawa coupling, forward-backward and charge asymmetries, spin correlations and polarization, and W boson polarization. The results are compared to the standard model predictions and limits on new physics from these measurements are also presented.}

\FullConference{%
  The Eighth Annual Conference on Large Hadron Collider Physics-LHCP2020 \\
  25-30 May, 2020\\
  online}

\begin{document}
\maketitle

%\linenumbers

\section{Introduction}
With a mass m$_\mathrm{t}\simeq$173 GeV, the top quark is the most massive elementary particle known to date. It has a very short lifetime ($\sim$0.5$\times10^{-24}$ s), which is shorter than the hadronization timescale ($\sim$2.6$\times10^{-24}$ s). Therefore, no hadronic top quark pair ($\ttbar$) bound states can form and the properties of top quarks are accessible experimentally. 
Moreover, its lifetime is shorter than the spin de-correlation time scale ($\sim$3$\times10^{-21}$ s). Hence, in $\ttbar$ production, top quark spins stay correlated and the spin correlations are measurable from the decay products. 
This proceeding focuses on the recent top quark properties measurements from ATLAS~\cite{atlas} and CMS~\cite{cms} at the LHC.

\section{Top Quark Mass and Width}
The mass of the top quark can be extracted from its decay products. 
The ATLAS and CMS measurements~\cite{atlas_mass_comb,cms_mass_comb}, each using the combined dataset collected at $\sqrt{s}=7$ and 8 TeV, have a precision of $\sim$500 MeV that corresponds to $\sim$0.28\%. These measurements are limited by jet energy scale (JES) calibration, b-tagging and modeling uncertainties. Many of the individual measurements have an uncertainty smaller than 1 GeV, and at this level of precision, the interpretation of m$_\mathrm{t}$ measurements is  complicated by non-perturbative effects (up to $\sim$1 GeV). Therefore, it is important to measure m$_\mathrm{t}$ in well-defined mass schemes and with independent methods resulting in different sources of systematic uncertainties. Alternative methods to extract m$_\mathrm{t}$ from production observables using total and differential $\ttbar$ cross sections have been made at Run I and Run II of the LHC. These measurements are dominated by $\ttbar$ threshold (m$_{\ttbar}\sim2\rm{m}_\mathrm{t}$) production where the uncertainties due to parton distribution functions (PDFs) and higher-order corrections become important. 

The most precise CMS measurement using production observables uses triple-differential normalized cross sections in bins of number of jets ($\rm{N_{jet}}^{0,1+}$), $\ttbar$ invariant mass (m$_{\ttbar}$) and rapidity (y$_{\ttbar}$)~\cite{cms_triple_diff_mass} to simultaneously extract PDF, the strong coupling parameter ($\alpha_s$) and m$_\mathrm{t}$ at next-to-leading order (NLO). The threshold region is the most sensitive region. The measured top quark pole mass and strong coupling parameter are m$_\mathrm{t}^{\rm {pole}}=170.5\pm0.8$ GeV and $\alpha_s=0.1135^{+0.0021}_{-0.0017}$. The precision of the top quark mass extraction is $\sim$0.5\% dominated by experimental and modeling uncertainties. Possible effects from soft-gluon and Coulomb corrections \cite{coulomb} near the threshold are not studied in detail because these effects are known only with large uncertainties in the cross section. 

The most precise ATLAS measurement using production observables is made analyzing $\ttbar$+1 jet events in the lepton+jets channel to extract m$_\mathrm{t}^{\rm{pole}}$ and the running mass in the modified minimal subtraction scheme ($\overline{\rm{MS}}$ or m$_\mathrm{t}(\rm{m}_\mathrm{t}))$~\cite{atlas_tt1j_mass}. In order to do this, the $\rho_s=2\rm{m_0}/\rm{m}_{\ttbar+{\rm jet}}$ variable is used, with the reference mass scale m$_0=170$ GeV. 
The parton level distribution of $\rho_s$ is compared with QCD NLO + parton shower (PS) calculations to extract the masses
    m$_\mathrm{t}(\rm{m}_\mathrm{t})=162.9\pm0.5(\rm{stat})\pm1.0(\rm{syst})^{+2.1}_{-1.2}(theo)$~GeV and
    m$_\mathrm{t}^{\rm{pole}}=171.1\pm0.4(\rm{stat})\pm0.9(\rm{syst})^{+0.7}_{-0.3}(theo)$~GeV, respectively.
These measurements are dominated by the scale, PDF, PS, color reconnection (CR), and JES uncertainties. The extracted $\overline{\rm MS}$ mass has a larger theory uncertainty and this is due to the larger dependence on scales at the production threshold.
It is also verified that m$_\mathrm{t}(\rm{m}_\mathrm{t})$ translated to m$_\mathrm{t}^{\rm{pole}}$ using the NLO QCD formula is in good agreement with the extracted value. 

 CMS made the first experimental investigation of the running of the top quark mass~\cite{cms_running}. 
 $\overline{\rm MS}$ is extracted at one loop precision as a function of m$_{\ttbar}$ by comparing NLO calculations to the $d\sigma(\ttbar)/d\rm{m}(\ttbar)$ measurement corrected to the parton level in the $\mathrm{e^\mp\mu^\pm}$ channel using 2016 data at $\sqrt{s}=13$~TeV.  
The extracted running of m$_\mathrm{t}$ up to $\sim$1 TeV is in agreement with the scale dependence predicted by the renormalization group equations evolved either from an initial scale of $\mu_0=\mu_{\rm{ref}}=476$~GeV or from an initial scale of $\mu_0=\rm{m}_{\mathrm t}^{\rm{inc}}=163$~GeV extracted at NLO from the inclusive $\sigma_{\ttbar}$ measurement. 
For the latter case, the measurement uncertainties (i.e. fit, extrapolation and PDF) are also evolved to higher scales. 
 
The top quark mass is also measured from the jet mass (m$_{\rm{jet}}$) distribution constructed from hadronic decay products of boosted top quarks by CMS in the lepton plus jets channel at $\sqrt{s}=13$ TeV~\cite{cms_boosted_mass}. For the measurement, the XCone algorithm~\cite{xcone} is used to reconstruct large-radius XCone jets with $\rm{p_T}>$ 400 GeV and subjets within the large-radius jets from which XCone m$_{\rm{jet}}$ is reconstructed. The unfolded m$_{\rm{jet}}$ distribution at particle level is used to extract  m$_\mathrm{t}=172.6\pm0.4(\rm{stat})\pm1.6(\rm{exp})\pm1.5(\rm{model})\pm1.0(\rm{theo})$~GeV with a precision of 1.4\%. The measurement is dominated by JES, XCone jet energy correction, final-state radiation, CR, and underlying event tune uncertainties. The average energy scale of the measurement is $\sim$480 GeV, a value significantly higher than the scale accessed in other m$_\mathrm{t}$ measurements. This measurement may help understand the ambiguities between Monte Carlo (MC) and m$_\mathrm{t}^{\rm{pole}}$, and also eventually enable a determination of m$_\mathrm{t}$ using analytical calculations which are only available in the boosted regime in the soft-collinear effective theory~\cite{hoang}.

The top quark mass is also measured using soft-muon tags by the ATLAS experiment~\cite{soft_atlas}. The soft muon from the B hadron and the muon from the W  boson from the top quark decay are used to reconstruct an invariant mass which is sensitive to m$_\mathrm{t}$. The lepton+jets channel is used with the requirement of two b-tagged jets, one with displaced vertex and one with the soft-muon tag. A simultaneous template fit to m$(\ell,\mu)$ distributions from same-sign and opposite-sign samples is perfomed to extract the result.
To increase the precision of the measurement, the fragmentation function used in PYTHIA8~\cite{pythia8} is improved using a new fit to LEP data. The b- and c-hadron decay branching fractions are also adjusted to match that of the previous measurements. The extracted mass is m$_\mathrm{t}=174.48\pm0.40(\rm{stat})\pm0.67(\rm{syst})$~GeV with a precision of $\sim$0.45\%. The measurement is sensitive to different modeling effects than those present in other top quark mass measurements, namely, heavy-flavor hadron decay modeling, pile-up, and b-quark fragmentation. This will make it useful in combinations with measurements with different modeling effects.

The top-quark decay width, $\Gamma_\mathrm{t}$, is measured by the ATLAS experiment directly using a profile likelihood template fit to m$(\ell,\rm{b})$ distribution in the dilepton channel using the full Run II data~\cite{atlas_top_width}. 
The measurement yields $\Gamma_\mathrm{t}=1.94^{+0.52}_{-0.49}$ GeV and is in good agreement with SM NNLO predictions. The dominant uncertainties of the measurement are jet reconstruction, signal and background modeling, limited size of the sample of simulated events, and flavor tagging.

\section{Top Quark Yukawa Coupling}
The Yukawa coupling is measured by CMS using full Run II data utilizing possible weak corrections from new physics that modify  differential cross sections~\cite{cms_yukawa}. The Yukawa parameter, Y$_\mathrm{t}$, defined as the Yukawa coupling strength normalized to the standard model (SM) Yukawa coupling strength is used for the measurement. Weak corrections as a function of m$_{\ttbar}$ and $\Delta$y$_{\ttbar}$ at various Y$_\mathrm{t}$ values are derived using the HATHOR code~\cite{hathor} and are applied to all $\ttbar$ samples so that their kinematic properties remain dependent on Y$_\mathrm{t}$. The measurement is done in the dilepton channel, requiring two opposite sign leptons and two b jets.  Partially-reconstructed kinematic variables  M$_{\mathrm{b\ell}}=m\mathrm{(b+\overline{b}+\ell+\overline{\ell})}$ and $|\Delta \rm{y}|_{\rm b\ell}=|\rm{y(b+\overline{\ell})-y(\overline{b}+\ell)}|$ are used for the measurement. For the latter variable each jet is matched to the correct lepton through a kinematic fit. The minimization of the log likelihood in the fit to data yields Y$_\mathrm{t}=1.16^{+0.24}_{-0.35}$ or [0.00,1.62] at 95\% CL. Fig.~1 displays the likelihood scan and data and MC events in M$_{\mathrm{b\ell}}$ bins for different data taking periods and  two $|\Delta \rm{y}|$ bins. The measurement is dominated by electroeak (EWK) corrections, PS and matrix element scales, and JES uncertainties. The measurement can be compared to the other CMS result that is exclusively dependent on the top Yukawa coupling made using four top quark production for which the upper limit at 95\% CL is 1.7~\cite{four_top}. The best, yet model-dependent measurement, of this parameter comes from the CMS Higgs combination~\cite{cms_higgs_yukawa}.

\begin{figure}
    \centering
    \includegraphics[width=6.0cm]{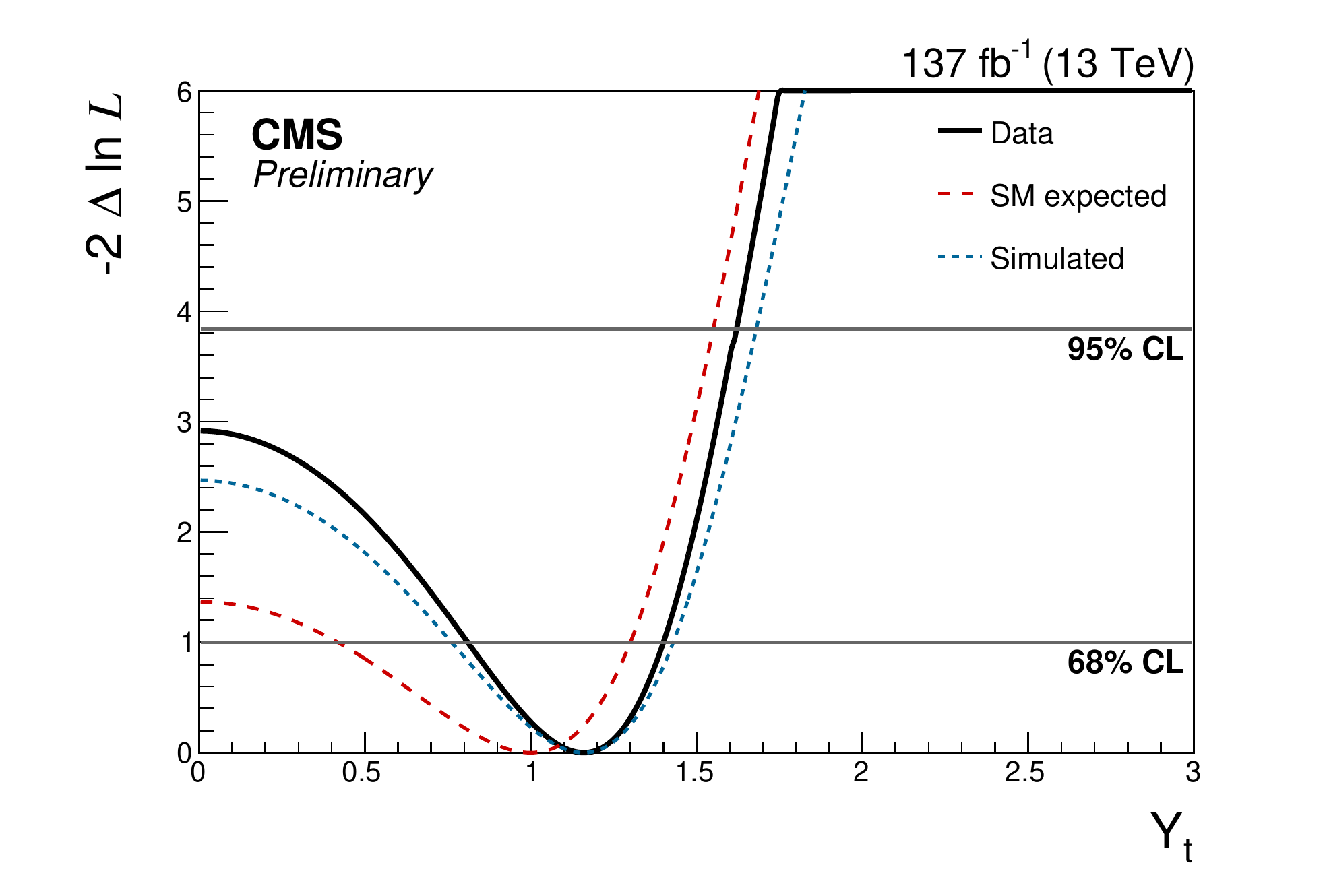}
    \includegraphics[width=5.6cm]{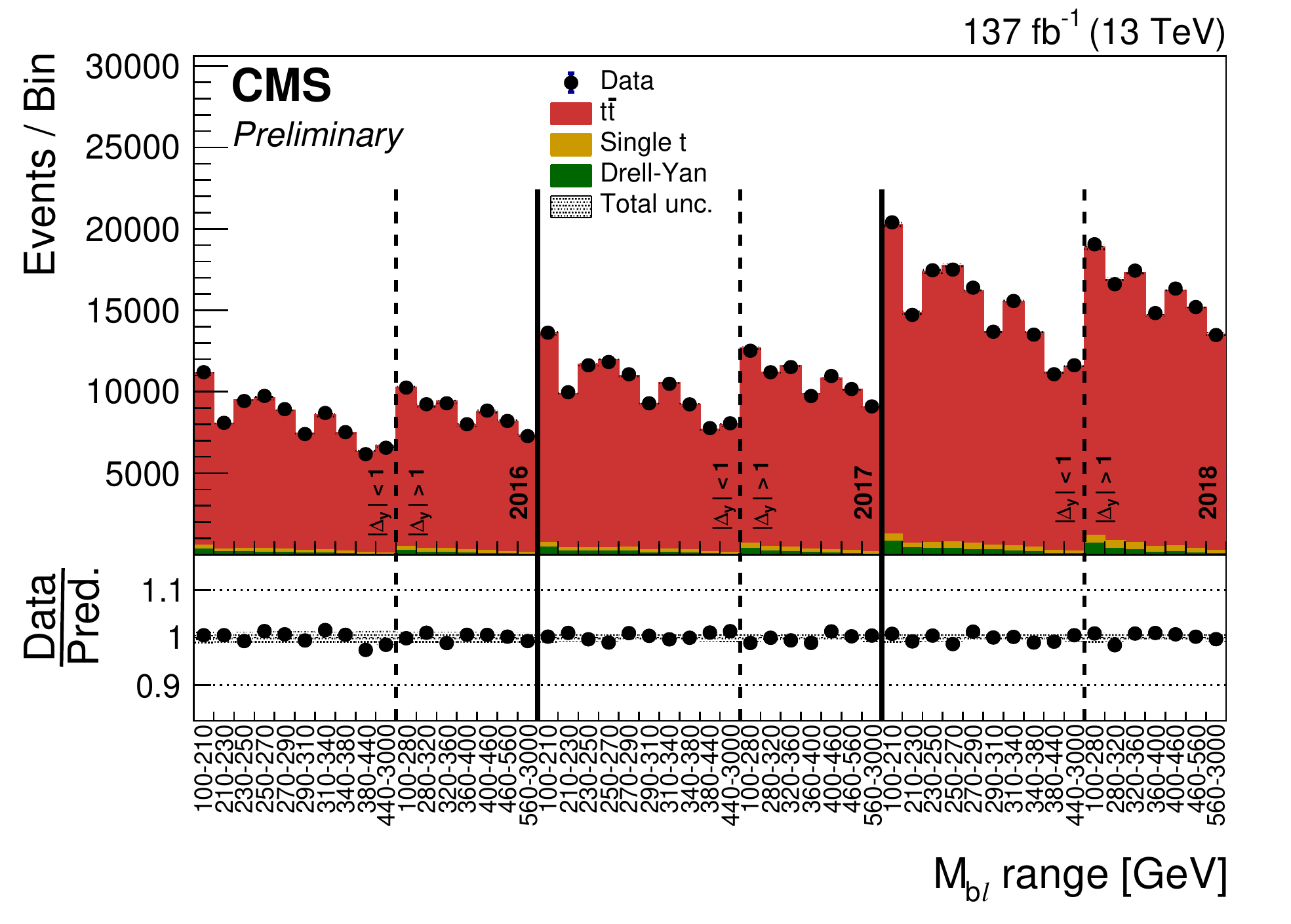}
    \caption{The likelihood scan (left). Expected curves assume the SM value $\rm{Y_t}$= 1.0 (red, dashed) and the final best-fit value of $\rm{Y_t}$= 1.16 (blue, dashed). 
    The data and MC simulation in bins of M$_{\mathrm{b\ell}}$ at the best fit value of $\rm{Y_t}$= 1.16, with shaded bands displaying the post-fit uncertainty. The solid lines divide the three data taking periods, while the dashed lines divide the two $|\Delta \rm{y}|$ bins in each data taking period \cite{cms_yukawa}. 
    }
\end{figure}

\section{CKM Matrix Elements of Third Generation Quarks}
CMS measured the CKM matrix elements from single top quark t-channel using processes directly sensitive to $|\rm{V_{tb}}|$, $|\rm{V_{td}}|$, and $|\rm{V_{ts}}|$ matrix elements in production and decay~\cite{cms_ckm}. The yields of different signals are extracted through a simultaneous fit to data in different event categories, 2 jet + 1 b-tag, 3 jet + 1 b-tag, and 3 jet + 2 b-tags, using multivariate discriminators to  separate  signal and backgrounds. CKM matrix elements are inferred from signal strengths $\mu_{\rm b}=\sigma_{\rm t-chan.}\times \rm{BR(meas)}/\sigma_{\rm t-chan.}\times \rm{BR(theo)}$. The signal strength from the fit is $\mu_{\rm{b}}=0.99\pm0.12$. If CKM unitarity of SM is taken for granted, then the following limits are obtained at 95\% CL: $|\rm{V_{tb}}|>0.970$, $|\rm{V_{td}}|^2+|\rm{V_{ts}}|^2<0.057$. If this assumption is dropped, the existence of additional quark families with $\rm{m}>\rm{m}_\mathrm{t}$ is assumed, and the partial width of each top quark decay is left free: $|\rm{V_{tb}}|=0.988\pm0.051$, $|\rm{V_{td}}|^2+|\rm{V_{ts}}|^2=0.06\pm0.06$. If all assumptions are dropped except the assumption that the contributions to $\Gamma_\mathrm{t}$ from the mixing of the three families are negligible: $|\rm{V_{tb}}|=0.988\pm0.024$,
$|\rm{V_{td}}|^2+|\rm{V_{ts}}|^2=0.06\pm0.06$, and $\Gamma_\mathrm{t}^{\rm obs}/\Gamma_\mathrm{t}=0.99\pm0.42$.
The results are dominated by modeling uncertainties. These results  are the first direct model independent measurements of the CKM matrix elements for the third generation quarks, and are the best determinations of these  parameters.

\section{Top Quark Pair Forward-Backward and Charge Asymmetries}
CMS made the first $\ttbar$ forward-backward asymmetry ($\rm{A_{FB}}^{\ttbar}$) measurement at the LHC in the lepton+jets final states with resolved and boosted topologies selected and reconstructed through a kinematic fit~\cite{cms_afb}.
$\rm{A_{FB}}^{\ttbar}$ is defined using the angle ($\theta^{*}$) of the top quark relative to the direction of the initial-state parton in the $\ttbar$ center-of-mass frame. 
The measurement is done through multi-dimensional template-based likelihood fits to $\cos\theta^{*}$, m$_{\ttbar}$, and $\rm{x_F=2p_L}/\sqrt{s}$ where $\rm{p_L}$ is the longitudinal momentum of the $\ttbar$ system in the laboratory frame. The templates are based on an approximate-NLO model for  $\mathrm{q\overline{q}}$ and $\mathrm{gg}$-initiated subprocesses.
The fits yield $\rm{A_{FB}}^{\ttbar}=0.048^{+0.095}_{-0.087}(\rm{stat})^{+0.020}_{-0.029}(\rm{syst})$, anomalous chromomagnetic moment $\hat{\mu}_\mathrm{t}$=$-0.024^{+0.013}_{-0.009}(\rm{stat})^{+0.016}_{-0.011}(\rm{syst})$ and  chromoelectric moment $|\mathrm{\hat{d}}_\mathrm{t}|<0.03$ at 95\% CL.
The dominant uncertainties are due to boosted e+jets trigger efficiency, JES, b-tag efficiency, b fragmentation and renormalization and factorization scales. 
 The measured $\rm{A_{FB}}^{\ttbar}$ agrees well with measurements at the Tevatron, NNLO QCD calculations \cite{mitov14} and the results on the chromo moments are consistent with the ones obtained from CMS spin correlation measurements in the dilepton channel. 

The ATLAS Collaboration obtained the first evidence of $\ttbar$ charge asymmetry ($\rm{A_C}^{\ttbar}$) at the LHC in the lepton+jets channel combining the resolved and boosted topologies using the full Run II data~\cite{atlas_ac}. 
$\rm{A_C}^{\ttbar}$ is defined with respect to $\Delta|\rm{y}|=|\rm{y}_{\mathrm t}|-|\rm{y}_{\overline{\mathrm t}}|$.
The asymmetry is corrected to the parton level through a Bayesian unfolding procedure. 
The inclusive asymmetry is measured to be $\rm{A_C}^{\ttbar}=0.0060\pm0.0015$ which differs from zero by 4$\sigma$. 
The dominant uncertainties sources are initial-state radiation in $\ttbar$, W+jets factorization and renormalization scales, W+jets normalization, and missing tranverse energy resolution.  
$\rm{A_C}^{\ttbar}$ is measured in bins of m$_{\ttbar}$ as well. All measurements are consistent with QCD NNLO + EWK NLO calculations \cite{czakon17} and POWHEGv2~\cite{powheg}+PYTHIA8 simulation, and limits on linear combinations of Wilson coefficients of dimension-6 EFT operators are placed. 

\section{Spin Correlation of Top Quark Pairs and Polarization}
CMS measured the coefficients of the spin density matrix that characterizes the spin dependence of $\ttbar$ production~\cite{cms_sc}. The coefficients are determined by  angular distributions unfolded to parton level in the dilepton channel. Laboratory frame asymmetries that can not be directly related to the spin density matrix coefficients are also measured. All distributions and extracted parameters are found to be in agreement with the SM predictions at (N)NLO and made with POWHEGv2+PYTHIA8 and MG5\_aMC+PYTHIA8 using the FxFx merging scheme~\cite{mg5,fxfx}. 

In the dilepton channel, ATLAS made inclusive $\Delta\phi(\ell^+,\ell^-)$, and $\Delta\eta(\ell^+,\ell^-)$, measurements and differential measurements of $\Delta\phi(\ell^+,\ell^-)$ in bins of m$(\ttbar)$ and extracted the spin correlation fraction ($\rm{f_{SM}}$) using templates that are fitted to the parton-level distributions using spin-correlated and -uncorrelated hypotheses \cite{atlas_sc}. The parameter $\rm{f_{SM}}$ in SM is equal to 1. ATLAS measured $\rm{f_{SM}}=1.249\pm0.024(\rm{stat})\pm0.061(\rm{syst})^{+0.067}_{-0.090}(theo)$ dominated by initial/final-state radiation, and scale setting uncertainties, which shows a difference of 2.2$\sigma$ from the SM prediction. This led to some alternative predictions with NLO QCD + weak coupling and also NLO QCD using an expansion of the differential distribution in powers of the couplings. 
The extracted value of $\rm{f_{SM}}$ using the NLO QCD + weak-expanded template is $\rm{f_{SM}}=1.03\pm0.07(\rm{stat})\pm^{+0.10}_{-0.14}(scale)$ which is consistent with the SM, POWHEGv2+PYTHIA8 and the NLO predictions \cite{behring19}. It is less consistent with the expanded-NNLO prediction. The comparisons of the unfolded $\Delta\phi$ data to these theory calculations are displayed in Fig.~2.  
The measurement of $\Delta\phi$ in bins of $\Delta\eta$ is also used to place limits on top squarks in the m=170-230 GeV range. 

\begin{figure}
    \centering
    \includegraphics[width=6.0cm]{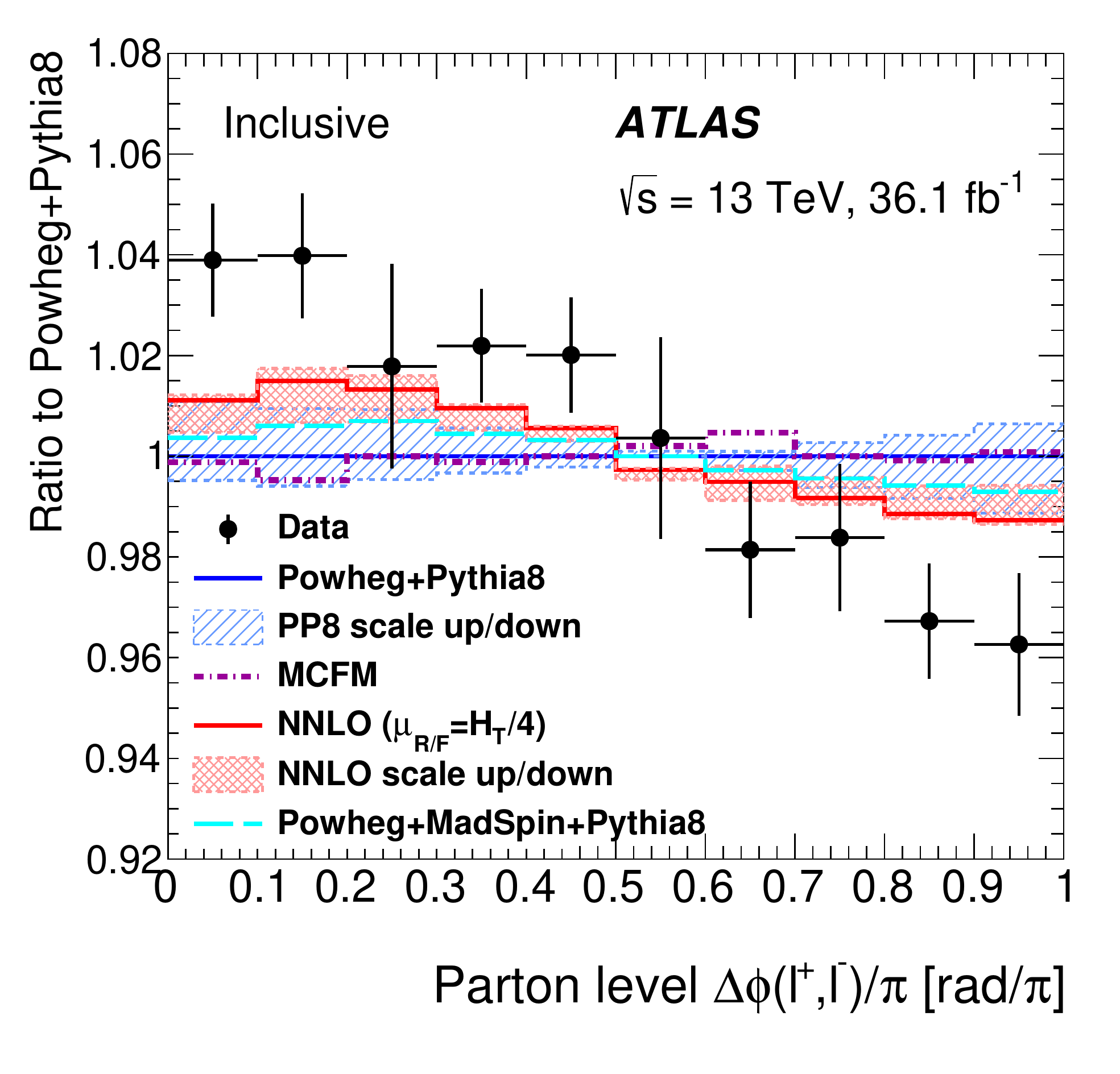}
    \includegraphics[width=6.0cm]{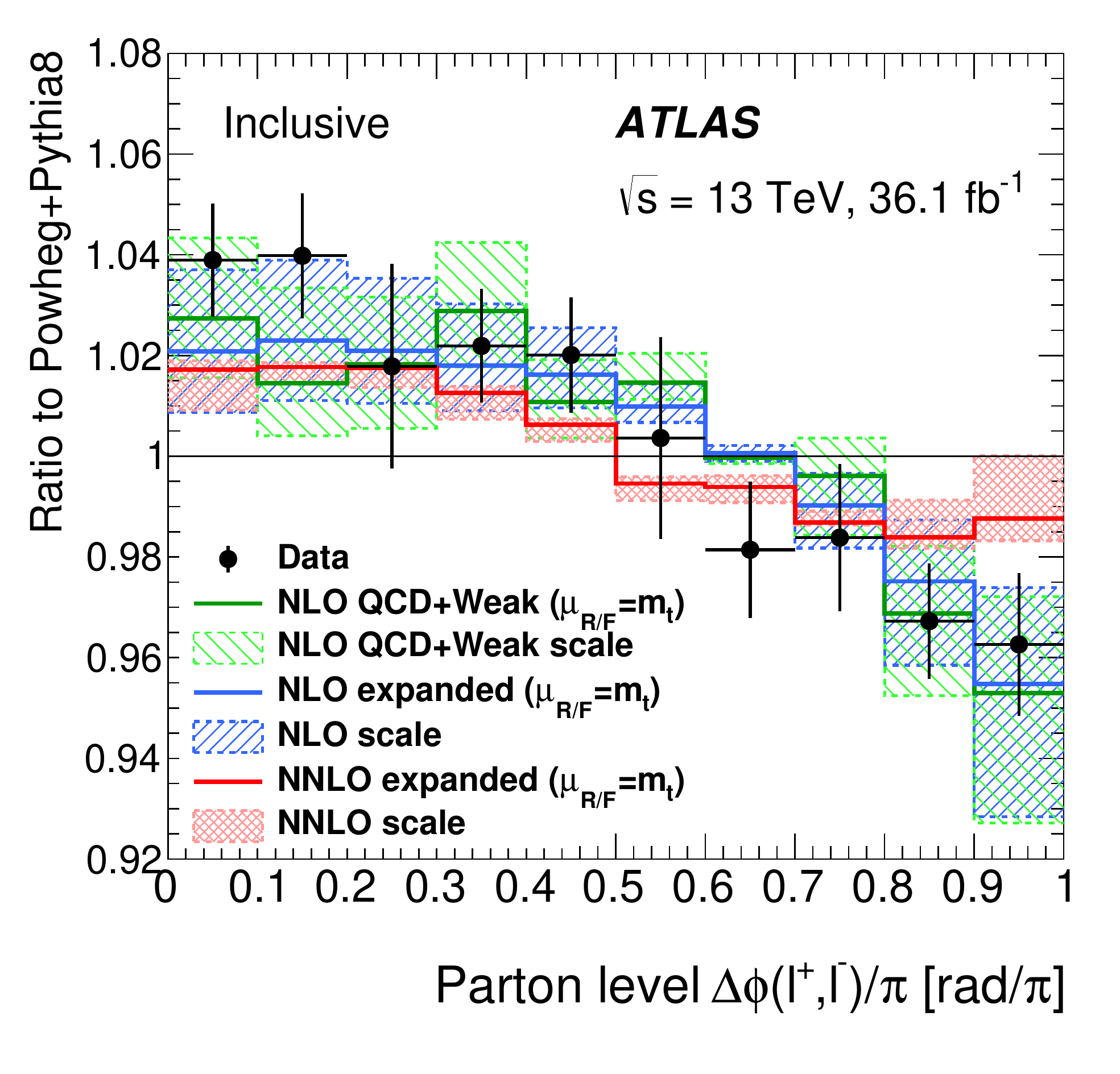}
    \caption{Ratios of unfolded data and theoretical predictions for $\Delta\phi$ for the inclusive selection. Comparisons to NLO generators and NNLO fixed-order predictions (left), and to NLO and NNLO calculations expanded in the normalized cross-section ratio (right) are displayed \cite{atlas_sc}. }
\end{figure}

The LHC Top Physics Working Group (LHC$top$WG) made comparisons of the ATLAS and CMS normalized cross sections in bins of $|\Delta\phi(\ell^+,\ell^-)|$ at the parton level~\cite{toplhcwg_sc}. Very good agreement between ATLAS and CMS data and between ATLAS and CMS main MC predictions is observed. A good agreement of data with MG5\_aMC with FxFx merging is also observed as well as a fair agreement with the NNLO calculation \cite{behring19}. These comparisons pave the way for the first $\sqrt{s}$=13 TeV ATLAS+CMS combination from the LHC$top$WG.  

\section{W Boson Polarization}
The LHC$top$WG made the first full combination of the $\sqrt{s}=8$ TeV data for W boson helicity fractions ($\rm{F_X}$) measured by ATLAS and CMS (with 20.2 and 19.7 fb$^{- 1}$, respectively) using both $\ttbar$ and single top measurements \cite{W_hel}. Several correlation schemes are probed in the combination and the resulting uncertainties cover the observed deviations. The results are dominated by statistical, background, radiation/scales uncertainties, and limited size of the sample of simulated events. The results agree with NNLO QCD calculations. The precision for longitudinal ($\rm{F_0}$) and left-handed ($\rm{F_L}$) polarization is $\sim$2\% and $\sim$3.5\%, respectively. The improvement in precision with respect to the most single precise measurement for $\rm{F_0}$ and $\rm{F_L}$ is 25\% and 29\%, respectively. With the combined data, confidence limits for anomalous couplings and Wilson coefficients are also derived.

\end{document}